\documentclass[conference]{IEEEtran}
\IEEEoverridecommandlockouts

\usepackage{amsmath,url}

\usepackage{cite}
\usepackage{amsmath,amssymb,amsfonts}
\usepackage{algorithmic}
\usepackage[]{graphicx}
\graphicspath{{Figures/}}
\usepackage{caption}
\usepackage{subfig}
\usepackage{lipsum}
\usepackage{balance}
\usepackage{textcomp}
\usepackage{xcolor}
\usepackage{multirow}
\usepackage{color}
\usepackage{xcolor}
\usepackage{algorithm} 
\usepackage{algorithmic}
\usepackage{listings}
\usepackage{courier}
\usepackage{float}

\definecolor{CustomColor}{RGB}{255, 255,255}

\lstdefinelanguage{turtle}
{
    columns=fullflexible,
    keywordstyle=\color{red},
    morekeywords={@prefix,@base,@forSome,@forAll,@keywords},
    morecomment=[l]{\#},
    tabsize=4, 
    alsoletter={-?}, 
    morecomment=[s][\color{blue}]{<}{>},
    basicstyle=\ttfamily\color{black}, 
    morestring=[b][\color{black}]\",    
    backgroundcolor=\color{CustomColor}
}
\lstdefinestyle{turtle}{%
    morekeywords={a, @prefix},
    morecomment=[s][\rmfamily]{<}{>},
    morecomment=[s][\itshape]{"}{"},
}

\def\BibTeX{{\rm B\kern-.05em{\sc i\kern-.025em b}\kern-.08em
    T\kern-.1667em\lower.7ex\hbox{E}\kern-.125emX}}

\usepackage{nomencl}
\makenomenclature

\begin{document}

\title{Interlinking Heterogeneous Data \\for Smart Energy Systems}


\author{
Fabrizio~Orlandi$^{1}$,
Alan~Meehan$^{1}$,
Murhaf~Hossari$^{1}$,
Soumyabrata~Dev$^{1}$,
Declan O'Sullivan$^{1}$,
and~Tarek~AlSkaif$^{2}$
\\
$^{1}$The ADAPT SFI Research Centre, Trinity College Dublin, Dublin, Ireland \\
$^{2}$~Copernicus Institute of Sustainable Development, Utrecht University, Utrecht, The Netherlands
\thanks{
This work is supported by the the Joint Programming Initiative (JPI) Urban Europe project: “PARticipatory platform for sustainable ENergy managemenT (PARENT)” and the Netherlands Science Foundation (NWO).
}
\thanks{The  ADAPT  Centre  for  Digital  Content  Technology  is  funded  under  the  SFI Research Centres Programme (Grant 13/RC/2106) and is co-funded under the European Regional Development Fund.}
\thanks{
Send correspondence to T. AlSkaif, E-mail:
t.a.alskaif@uu.nl
}
}

\IEEEoverridecommandlockouts
\IEEEpubid{\makebox[\columnwidth]{978-1-7281-1156-8/19/\$31.00~
\copyright2019
IEEE \hfill} \hspace{\columnsep}\makebox[\columnwidth]{ }} 

\maketitle


\begin{abstract}
Smart energy systems in general, and solar energy analysis in particular, have recently gained increasing interest. This is mainly due to stronger focus on smart energy saving solutions and recent developments in photovoltaic (PV) cells. Various data-driven and machine-learning frameworks are being proposed by the research community. However, these frameworks perform their analysis - and are designed on - specific, heterogeneous and isolated datasets, distributed across different sites and sources, making it hard to compare results and reproduce the analysis on similar data. 
We propose an approach based on Web (W3C) standards and Linked Data technologies for representing and converting PV and weather records into an Resource Description Framework (RDF) graph-based data format. This format, and the presented approach, is ideal in a data integration scenario where data needs to be converted into homogeneous form and different datasets could be interlinked for distributed analysis. 
\end{abstract}

\nomenclature{API}{Application Programming Interface}
\nomenclature{CSV}{Comma-Separated Values}
\nomenclature{DB}{Database}
\nomenclature{RDB}{Relational Database}
\nomenclature{JSON}{JavaScript Object Notation}
\nomenclature{PV}{Photovoltaic}
\nomenclature{KG}{Knowledge Graph}
\nomenclature{PWA}{Photovoltaic and Weather Analysis}
\nomenclature{R2RML}{RDB to RDF Mapping Language}
\nomenclature{RDF}{Resource Description Framework}
\nomenclature{SPARQL}{SPARQL Protocol and RDF Query Language}
\nomenclature{SSN}{Semantic Sensor Network Ontology}
\nomenclature{W3C}{World Wide Web Consortium (Web standards and Linked Data technologies)}

\printnomenclature

\section{Introduction}
\label{sec:1}
With the recent developments in photovoltaic (PV) cells, there has been a renewed interest in the area of solar energy generation and forecasting. Most of the work in solar analytics involves mining data and proposing data-driven, machine-learning frameworks. These data are of diverse types, and are generated from various sensors~\cite{manandhar2018analyzing, manandhar2018data, manandhar2018systematic}. Also such data comes in many forms (e.g. geological, medical, census, weather) and formats (e.g. RDB, CSV, JSON, APIs). 

\subsection{Motivation \& Background}
It is important to systematically analyse the diverse data types, such that we can identify trends, make predictions and inform decision-makers. Unfortunately, data is usually distributed across different sites and sources, and requires increasing amount of manual effort in discovering relevant pieces of information and pre-processing it. We can further unlock the potential of data if it is `linked' and `machine-readable'. The Web is moving away from information that is purely for human consumption, and instead expanding to include machine-readable data. This machine-readable data is being published in a format that allows computers to automatically understand how different pieces of data are connected to one another. This data\footnote{https://lod-cloud.net/} is known as \emph{Linked Data}~\cite{bizer2011linked}. 

\subsection{Relevant Literature}
Linked Data follows a decade of research in the Semantic Web domain~\cite{berners2001semantic} and has recently reached considerable popularity under the name of `knowledge graphs' (KGs)~\cite{bonatti_et_al:DR:2019:10328}. KGs have gained increasing popularity over the last years, especially in industry where they are now at the core of relevant consumer products (e.g. Google\footnote{\url{https://developers.google.com/knowledge-graph/}} and Bing\footnote{\url{https://azure.microsoft.com/en-us/services/cognitive-services/bing-entity-search-api/}} search engines). According to a recent business report~\cite{forrester}, ``51 percent of global data and analytics technology decision makers are either implementing, have already implemented, or are upgrading their graph database''. 
KGs are knowledge-bases of facts about entities and concepts (e.g., places, persons, artifacts) which are represented using the flexible structure of a graph. Facts are often extracted from encyclopedic knowledge, such as Wikipedia, or existing structured repositories (e.g. Wikidata\footnote{\url{https://www.wikidata.org/wiki/Wikidata:Main_Page}}), or even from unstructured sources such as social media posts (e.g. Facebook Graph\footnote{\url{https://developers.facebook.com/docs/graph-api}}). More details of the relevant literature can be found in Section~\ref{sec:RW}.

\subsection{Contributions \& Organization of the paper}
Our contribution in this paper is two fold: (a) we propose an RDF ontology for the modelling and description of photovoltaic records with a weather component; and (b) we demonstrate a process for the conversion and storage of disparate photovoltaic and weather data to RDF data. The RDF, graph-based, data format is ideal in a data integration scenario where data, of multiple formats, is to be converted into a homogeneous form. A predefined schema (tables, columns, relationships, constraints) does not need to be created in this case, instead dynamic data records can be created on the fly. These data records can then simply be linked together to form an intricate knowledge graph. Our Photovoltaic and Weather Analysis (PWA) ontology adds the semantic layer to this knowledge graph by supplying the terms which describe the data (the photovoltaic and weather data), thus adding semantics and increasing the machine readability of the data. Our data conversion process outlines the steps involved to transform the photovoltaic and weather data to RDF, and store that data to allow its retrieval and analysis using the SPARQL query language.

The remainder of this paper is structured as follows: \textit{Section~\ref{sec:RW}} presents related work and background in the areas of Linked Data, ontologies and techniques used for the conversion of non-RDF data to RDF - also known as semantic uplift; In \textit{Section~\ref{sec:PWA}} we detail our PWA ontology and the process we propose to semantically uplift data according to this ontology; \textit{Section~\ref{sec:UC}} presents possible use cases for our approach; \textit{Section~\ref{sec:Conc}} finishes the paper with conclusions and future work.

\section{Related Work}
\label{sec:RW}

\subsection{Linked Data}
Data that is made available on the Web in `Linked Data' format~\cite{bizer2011linked}, has great potential in that it can be easily integrated, published and connected through interlinks and Web (W3C) standards\footnote{\url{https://www.w3.org/DesignIssues/LinkedData}}. Widespread adoption of Linked Data is changing the way we use the Internet, and dramatically enhance decision-making across all sectors. Linked Data has multiple benefits\footnote{\url{https://www.w3.org/standards/semanticweb/data.html}}, for example:

\begin{itemize}
    \item Data will no longer be stored in static pieces in different places all over the Web. Instead, it will be connected and will provide greater context and a deeper understanding of data relationships (thanks to ontologies that define the schema and the meaning of data)~\cite{noy2004semantic}.
    \item Data will be machine-readable, so that computers can do more `thinking' on people's behalf by using logical inference and reasoning~\cite{domingue2011handbook}.
    \item Data will always be up-to-date and live federated queries can be performed online in real-time. Data is simply updated at the sources and Web (W3C) standards, such as SPARQL, allow for advanced SQL-like querying capabilities on the Web across different sources~\cite{harris2013sparql}.
    \item It greatly reduces time and costs for discovering, pre-processing and integrating data~\cite{manola2004rdf}.
\end{itemize}

\subsection{Ontologies}

A wide range of `ontologies', alternatively called `vocabularies', have been created in order to represent knowledge, entities and concepts for different domains~\cite{vandenbussche2017linked}. These vocabularies define the schema of semantic KGs and describe their entities and relationships unambiguously. Ontologies are typically designed to be published online, reused and eventually extended\footnote{\url{https://lov.linkeddata.es/dataset/lov/}}.


In the domain of weather and renewable energy a few vocabularies exist and can be reused for our purpose.
The most relevant one would be the AEMET ontology\footnote{\url{http://aemet.linkeddata.es/ontology/}}, a meteorology ontology network that extends the W3C Semantic Sensor Network Ontology (SSN)\footnote{\url{https://www.w3.org/TR/vocab-ssn/}}. The aim of this ontology network is
to represent knowledge related to measurements made
by weather stations. The AEMET project itself, aims at making data sources from the Spanish Meteorological Office available as Linked Data~\cite{atemezing2013transforming}. The project collects data from approximately 250 automatic weather stations deployed across Spain and available as CSV files. These files are transformed (or ``uplifted'') to RDF data according to Linked Data principles and the AEMET ontology. 
While this ontology is mainly focused on weather terms and concepts, the aforementioned SSN ontology~\cite{compton2012ssn} is more generic and can be used as an extensible core ontology for representing any sensor observation. 
In this work, we take inspiration from these semantic models specifically designed for weather and sensor data and derive our own model tailored at integrating weather data with solar and renewable energy data. 

In addition to the semantic models (or ontologies), a software architecture for continuously extracting and transforming this data is proposed in this paper. Strategies for making this data available online following W3C standards are described as well as use cases showing advanced querying capabilities and interlinking with different datasets (e.g. interlinking weather data with energy data or geographical information). 
We build on existing research, such as~\cite{Reddy2019}~\cite{radulovic2015guidelines} and~\cite{calbimonte2012deriving}. In~\cite{Reddy2019} the authors integrate weather information together with health records using Linked Data principles and a conversion process similar to ours. In~\cite{radulovic2015guidelines} the authors propose guidelines for Linked Data generation and publication of building energy consumption. The SSN ontology is reused for their purpose, as data from various sensors needs to be captured. Instead, in~\cite{calbimonte2012deriving} examples of data analysis on semantic sensor metadata modelled using the SSN ontology are described.
In our work, we designed a core ontology (that can be aligned to the SSN and AEMET ontologies) which specifically targets the integration of weather data with PV energy data and that eases the process of mapping and uplifting common datasets used by researchers in the domain.

\begin{figure*}[h]
  \centering
	\includegraphics[width=0.8\textwidth]{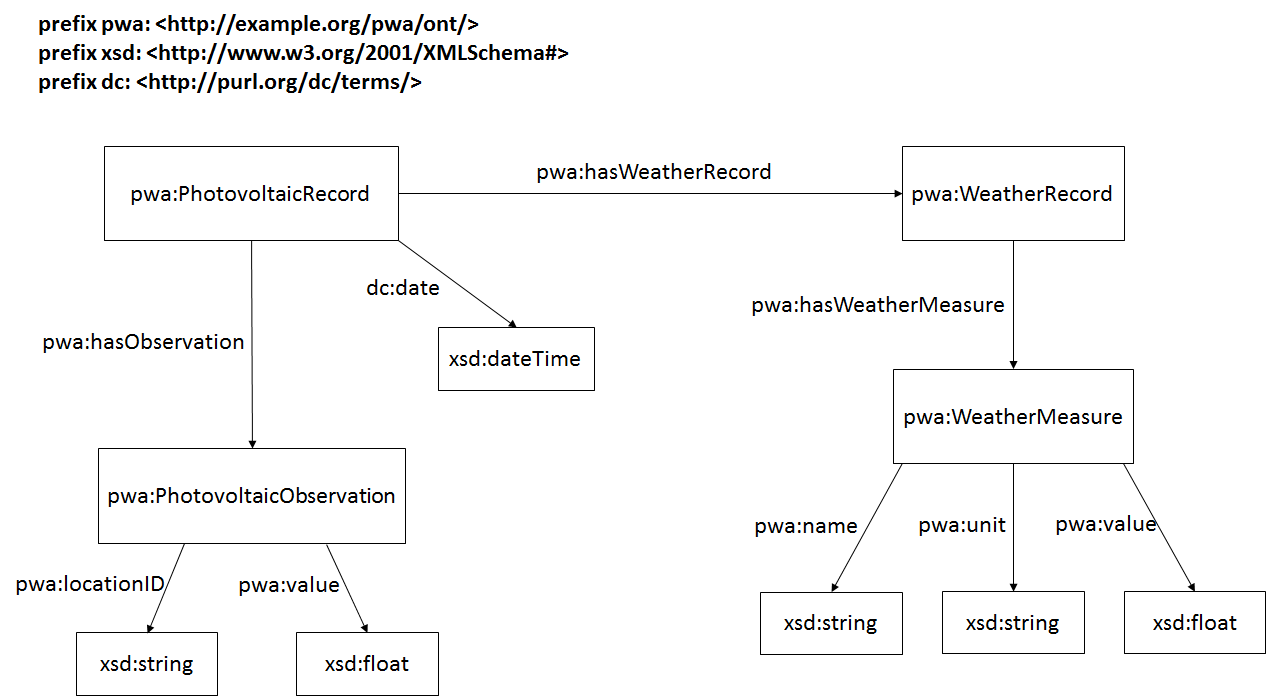}
  \caption{The Photovoltaic and Weather Analysis (PWA) Ontology}
	\label{fig:pwa_ont}
\end{figure*}

\subsection{Semantic Uplift}
The term semantic uplift~\cite{crotti01} is used to denote the process of converting data of a non-RDF format into RDF according to a particular ontology and is very much a classical data mapping process~\cite{kalfoglou2003ontology}. The purpose of such a process is to improve interoperability across data sets through transformation of data to a single format (RDF) and the addition of semantic meaning to the data - further enabling actions such as semantic search and improving the overall machine readability of the data. Semantic Uplift can be performed in many different ways, from custom scripts that perform the action as a once off task to specialised mappings languages that are used to create declarative mappings. Here we will briefly cover three approaches for semantic uplift.

The RDB to RDF Mapping Language (R2RML)~\cite{das2016r2rml} is a W3C recommendation for the conversion of relational and tabular data to RDF - through the creation of declarative mappings. R2RML is an expressive language that can be used to both structure the source data and describe it according to an RDF ontology. R2RML is becoming a mature standard at this stage and is already supported by some existing applications such as the Oracle and Stardog databases.

The RDF Mapping Language (RML) ~\cite{dimou2014rml} is an extension to R2RML which supports the conversion of an increased number of data formats (Relational, CSV, TSV, JSON and XML) to RDF. RML is also a declarative mapping approach with just as much expressivity as R2RML, however, not being a W3C standard, it is less well known and it does not have the same level of support in existing applications compared to R2RML.

SPARQL-generate ~\cite{lefranccois2017sparql} is a declarative mapping approach to convert a multitute of data formats (Relational, CSV, TSV, JSON, XML, EXI and CBOR) into RDF. The mappings are written in a syntax that is similar to the SPARQL language, but extended with additional clauses to provide the functionality required of the mapping language. Like RML, SPARQL-generate is not a W3C standard, therefore there are less tools and applications which provide support for it.

In our data conversion process that we describe in the next section, we utilise R2RML as it is the most mature of the existing approaches.

\section{PWA Ontology and Data Conversion Process}
\label{sec:PWA}
This section presents the ontology we developed to represent photovoltaic and weather data as well as the process we propose for the conversion of the original source data to RDF. 

\subsection{The Ontology}
We propose a new RDF ontology, the Photovoltaic and Weather Analysis (PWA) ontology, that is specifically designed to model and describe PV data and associated weather data. The idea behind capturing both of these is that analysing PV data and weather data at a certain time and location will lead to insights into optimal operating conditions for specific PV cells and forecasting to detect these optimal conditions. The PWA ontology consists of four classes and seven properties (see Figure~\ref{fig:pwa_ont} for a visual representation of the ontology). 

The \textit{PhotovoltaicRecord} class is the main class of interest in the PWA ontology. It has a date attached to it via the \textit{dc:date} property; it can have an arbitrary number of \textit{PhotovoltaicObservations} attached to it by the \textit{hasObservation} property; and a \textit{WeatherRecord} attached via the \textit{hasWeatherRecord} property. Note that while each photovoltaic record can have an arbitrary number of observations, each of those observations should be within close proximity of each other since there is only one weather record per photovoltaic record. It does not make sense to have two observations if those locations are 30km apart as the weather is likely to be different in the different locations. We leave it up to the end user do determine the suitable distance between observations that will be part of one photovoltaic record.

The \textit{PhotovoltaicObservation} class captures the energy created by a particular PV cell. Attached to each observation is the location identification number of a cell by the \textit{locationID} property and the value of the energy generated by the cell via the \textit{value} property.

The \textit{WeatherRecord} class is used to model weather records which can have multiple different types of measures (such as temperature, humidity, cloud cover, wind speed, wind direction etc.). Each weather record can have an arbitrary number of \textit{WeatherMeasures} attached to it via the \textit{hasWeatherMeasure} property.

The \textit{WeatherMeasure} class is used to model individual weather measures. The name of the measure (e.g. temperature) is attached via the \textit{name} property; the unit of measurement of the measure (e.g. Fahrenheit/Celsius) is attached by the \textit{unit} property; and the value of the measure is attached by the \text{value} property.

\begin{figure*}[h]
  \centering
  \includegraphics[width=0.7\textwidth]{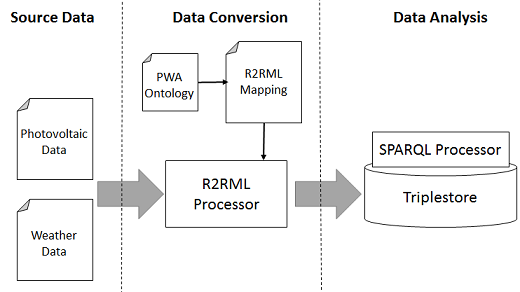}
  \caption{Data Conversion Process}
  \label{fig:process}
\end{figure*}

\begin{lstlisting}[style=turtle, caption={Example instance data described according to the PWA ontology.}, label={lst:turtle}]
01 @prefix pwa: <http://example.org/pwa/ont/>.
02 @prefix xsd: <http://www.w3.org/2001/XMLSchema#>.
03 @prefix dc: <http://purl.org/dc/terms/>.
04 @prefix data: <http://example.org/data/>.
05
06 data:1 a pwa:PhotovoltaicRecord; 
07   dc:date '2014-02-02T00:00:00'^^xsd:dateTime;
08   pwa:hasWeatherRecord 
            data:wr_2014-02-02T00:00:00;
09   pwa:hasObservation _:o1.
10
11 data:wr_2014-02-02T00:00:00 a pwa:WeatherRecord;
12   pwa:hasWeatherMeasure _:w1.
13
14 _:w1 a pwa:WeatherMeasure;
15   pwa:name 'cloud cover'^^xsd:string;
16   pwa:unit '%'^^xsd:string;
17   pwa:value '10'^^xsd:float.
18
19 _:o1 a pwa:PhotovoltaicObservation;
20   pwa:locationID '385'^^xsd:string;
21   pwa:value '0.02079'^^xsd:float.
\end{lstlisting}

An example of data modelled according to the PWA ontology is presented in Listing~\ref{lst:turtle}, encoded in RDF Turtle syntax. This example contains a photovoltaic record with simply one observation and the weather record also simply has one measure.

\subsection{The Conversion Process}
In our conversion process, displayed in Figure~\ref{fig:process}, we use R2RML as the mapping language for the conversion of relational and tabular data to RDF. We choose R2RML due to its expressivity and maturity level and also because it is a declarative mapping approach. This means there is a single mapping for each piece of data to be transformed. The idea is that by taking a declarative mapping approach, maintaining the data overtime is less troublesome if the original data sources change. Since there is a mapping for each data source, it means that if a data source has changed - you simply need to find the respective mapping and update it. This way, the mapping only needs to be changed and the source code does not need to be edited and recompiled each time a data source changes.

In our conversion process - we first analyse the source data sets to see which parts of that data \textit{aligns} with the classes and properties of the PWA ontology. When that is complete, The mappings are created which specify these alignments and the structure that the RDF is to adhere to. The R2RML mappings are then executed by the R2RML processor which takes as input the source data and the mappings and outputs the resulting RDF data. From there, this newly created RDF data is inserted into a triple-store, which is an RDF graph database, where it can now be accessed, queried and manipulated using the SPARQL query language.

\section{Use Cases}
\label{sec:UC}
As renewable energy is becoming more and more relied upon, energy providers will further have to balance the use of renewable energy sources against traditional energy generation (coal, oil, natural gas etc.) in order to meet the demand of the electricity grid. It is envisaged that the analyses of PV data and weather data will lead to more insight regarding the electricity being generated by these cells. Through these insights, more informed decisions could be made by energy providers as to whether how heavily they can rely on the renewable energy source and cut back on the traditional energy generation - saving natural resources and reducing the production of CO2 gas. With our PWA ontology, both PV and weather data are modelled and described in a homogeneous way, facilitating greater analysis of the overall data.

Modelling instance data according to the PWA ontology will allow, through querying, some interesting retrieval and analysis use cases. Consider, for instance, the following request to an information system: \textit{Retrieve the weather measures of all photovoltaic records where the average energy generated from the record's observations is greater than that of other photovoltaic records which have a lower cloud cover measure.} Having a system which is capable of providing an answer to such queries on top of distributed datasets may provide insight into the weather measures, other than cloud cover, which contribute to energy generation by the cells.

Another interesting retrieval use case would be, for example: \textit{Retrieve all photovoltaic records, their energy generated and cloud cover from associated weather records where the energy generated is greater than X and the cloud cover is greater than Y} (where \textit{X} and \textit{Y} are two numerical values). This request has been identified by domain experts in our team as potentially relevant for the data under consideration. It serves as another example of a use case for which our proposed solution would be beneficial. We display the SPARQL query to perform this use case in Listing~\ref{lst:SPARQL}. These are just two examples of data retrieval use cases than could be posed over the data. To note that these queries could be run against multiple federated data sources in real-time and would automatically aggregate the results. This is achievable simply using Linked Data standards and the proposed data modelling solution.

\begin{lstlisting}[style=turtle, caption={SPARQL query retrieving photovoltaic records with specific ``cloud cover'' and PV values (0.05 and 0.02 respectively)}, label={lst:SPARQL}]
01 PREFIX pwa: <http://example.org/pwa/ont/>
02 PREFIX xsd: <http://www.w3.org/2001/XMLSchema#>
03
04 SELECT ?PVRecord ?CloudCoverValue ?PVValue
05 WHERE 
06 {
07   ?PVRecord a pwa:PhotovoltaicRecord ; 
08         pwa:hasWeatherRecord ?WeatherRec ;
09         pwa:hasObservation ?PVObservation . 
10
11   ?WeatherRec pwa:hasWeatherMeasure ?WMeasure .  
12
13   ?WMeasure pwa:name "cloud cover" ;
14         pwa:value ?CloudCoverValue . 
15   FILTER (?CloudCoverValue > 0.05) .
16
17   ?PVObservation a pwa:PhotovoltaicObservation ;
18         pwa:value ?PVValue .
19   FILTER (?PVValue > 0.02) .
20 }
21 ORDER BY ?CloudCoverValue
\end{lstlisting}

\section{Conclusion and Future Work}
\label{sec:Conc}
In this paper we proposed an approach for representing and converting photovoltaic and weather records into Linked Data. The goal is to make such data interoperable and homogeneous so that it can be easily harmonised and analysed across heterogeneous data sources. This would allow researchers to publish data in an interoperable format and to execute their analysis on different datasets.
We proposed an RDF ontology, called PWA, for modelling and describing PV records with a weather component. Moreover, we described a process (based on R2RML) for the conversion, storage and querying of such data.
In order to show the potential of the proposed solution, two concrete use cases have been detailed.

As part of our future work, we will continue working on our identified use cases (cf.\ Section~\ref{sec:UC}). We plan to collect a large amount of data and publish it online following the described process and Linked Data principles. This data will be made available to researchers for online queries or as downloadable data dumps. Advanced queries and analysis algorithms will be implemented on top of this knowledge graph in order to test scalability, performance and compliance to requirements.
Alignment with different existing ontologies and extension of the model to additional types of data will also be investigated. 

\balance 

\bibliographystyle{IEEEbib}


\end{document}